\documentclass[11pt,a4paper]{article}
\usepackage{jcappub}
\usepackage{graphicx}

\title{Search for dark Higgs inflaton with curvature couplings at LHC experiments}

\author{Lucia Aurelia Popa}

\affiliation{Institute of Space Science,\\
 Bucharest-Magurele, Ro-077125 Romania}
 
\emailAdd{lpopa@spacescience.ro}

\abstract{ We analyse the dark Higgs inflation model with curvature corrections
and explore the possibility to test its predictions by the particle physics experiments at LHC.\\
We show that the dark Higgs inflation model with curvature corrections 
is  strongly favoured by  the present cosmological observation. 
The cosmological predictions of this model, including the quantum corrections of dark Higgs coupling constants and the uncertainty in estimation of the reheating temperature, lead to the dark Higgs mass $m_{\phi}=0.919 \pm$ 0.211 GeV  and the mixing angle 
$\theta =1.492 \pm$ 0.045 (at 68\% CL).\\
We evaluate the FASER and MAPP-1 experiments reach for dark Higgs inflaton mass and mixing angle in the 95\% CL cosmological confidence region for an integrated luminosity of 3ab$^{-1}$ at 13 TeV LHC, assuming 100\% detection efficiency. \\
We conclude that the  dark Higgs inflation model with curvature corrections 
is a compelling inflation scenario based on particle physics theory 
favoured by the present cosmological measurements that leaves imprints in the dark Higgs boson searchers at LHC.}

\def\be{\begin{equation}}
\def\ee{\end{equation}}
\def\ba{\begin{eqnarray}}
\def\ea{\end{eqnarray}}
\usepackage{amssymb,graphicx,subfigure}
\begin{document}
\begin{flushright}
\end{flushright}

\maketitle

\section{Introduction} 

The precise  Cosmic Microwave Background (CMB) properties 
reported by the {\sc Planck} experiment \cite{Planck_params,Planck_infl,BK15} and 
the discovery by LHC of the Higgs boson \cite{Atlas,CMS} 
 increased the interest in so called Higgs portal  interactions that connect 
 the  hidden (dark) sector and the  visible sector of the Standard Model (SM), 
 with expected imprints on collider experiments \cite{PBC}. 
 Scenarios beyond-the-SM (BSM), that introduce a dark sector in addition to the visible SM sector 
 are required to explain a number of observed phenomena in particle physics,
astrophysics and cosmology such as the non-zero neutrino masses and oscillations, the Dark Matter (DM), baryon asymmetry of the universe, the cosmological inflation.

It is usual to assume that cosmic inflation is decoupled from the SM at
energies lower than the inflationary scale since the slow-roll conditions for inflation generally permit only tiny
couplings of the inflaton field to other fields. This assumption  prevents
the direct investigation of inflation mechanism in particle physics experiments. Consequently,
 there are little compelling scenarios of inflation  based on particle physics theory.

Since the only known fundamental scalar quantum
field is the SM Higgs field, the inflation models using the  SM Higgs boson  
as inflaton attained great attention over the past years. 
A number of Higgs inflation models, mostly with non-canonical action, have been proposed. 
They include models with Higgs scalar field non-minimaly coupled to gravity
\cite{Bezrukov08,Futamase89,Fakir90}, non-minimal derivative coupling to the Einstein tensor 
\cite{Germani10,Granda11,Jimenez20,Tsu12}, 
scalar-tensor models \cite{Jimenez19a,Jimenez19b}, 
Galileon models \cite{Kamada11,Ohashi10,Kobayashi10,Kobayashi11}, quartic hilltop models \cite{Bramante16,German21}. \\
The viability of these models is already substantially limited
mostly because they predict tensor-to-scalar  ratios larger than the upper bound
set by the combined analysis  of {\sc Planck} and BICEP-Keck Array data (hereafter {\sc Planck}+BK15) that constrain 
the energy scale of inflation to \cite{Planck_infl,BK15}:
\begin{equation}
\label{infl_scale}
{V}^{1/4}_*=\left({\frac{3 \pi^2 A_s^{*}  }{2} r_{*}}\right)^{1/4} M_{pl} 
< 1.6 \times 10^{16} {\rm GeV}  \hspace{0.2cm}  (95\% {\rm CL}) \,.
\end{equation}
Here the quantities with $(^*)$ are evaluated at the pivot scale $k_*=0.002$, 
$r_*$ is the ratio of tensor-to-scalar amplitudes, $A_s^{*}$
is the amplitude of the curvature density perturbations and 
$M_{pl}  $  is the reduced Planck mass. This imply an upper bound for the Hubble expansion rate during inflation:
\begin{equation}
H_* < 2.5 \times 10^{-5} M_{pl}\, \hspace{0.2cm} (95\% {\rm CL}) \,.
\end{equation} 
The above bound selects  the  viable Higgs inflation models from the requirement  $H_{*} \ll \Lambda_c$,
where $\Lambda_c$ is the unitary bound of each underlying theory, defined as the scale below which the quantum gravitational corrections are sub-leading \cite{Burges09,Bezrukov11,Burges14}. \\
It worths to mention that  the chaotic inflation model with quartic potential is excluded  by the data at more than 
95\% confidence level \cite{Linde}.

Among the models used to lower the predictions for tensor-to-scalar ratio,
the most studied is the Higgs inflation with non-minimal coupling to gravity \cite{Bezrukov08}.
At tree level and for large non-minimal coupling   $\xi \sim {\cal O} (10^4)$, this model 
gives a small tensor-to-scalar ratio, 
 in agreement with the 
 {\sc Planck}+BK15 data.
 However, for such large values 
 of $\xi$ the unitary bound scale, $\Lambda_c=M_{pl}/\xi$, could be close or below the
energy scale of inflation \cite{Burges09,Calmet11}.  \\
An interesting framework for Higgs inflation is provided the scalar-tensor models including the
non-minimal kinetic coupling to the Einstein tensor and to the Gauss-Bonnet invariant.
These models can produce inflation simultaneously satisfying the present inflationary 
observational constrains and the unitary bound constraints \cite{Jimenez19a,Jimenez19b}.

Higgs portal interactions via the Renormalisation Group (RG) loop contributions can also  lower the
predictions of Higgs inflation models for the tensor-to-scalar ratio. 
The price to pay in these models  is the  electroweak (EW) vacuum metastability issue.
The actual values of Higgs boson and top quark masses imply
that the EW vacuum is metastable at energies larger than $\Lambda_I \sim 10^{11}$ GeV, 
where Higgs quartic coupling turns negative\footnote{The actual value of EW vacuum metastability scale is defined for the top quark mass $m_t=173.15$ GeV and Higgs boson mass $m_H=125.10$ GeV \cite{PDG} as the value of the Higgs field at which the Higgs quartic coupling, $\lambda_h$, becomes negative due to radiative corrections.}\cite{Bezrukov15,Buttazzo13,Degrassi12,Allison14}. \\
However, it is found that a small admixture of the Higgs field with a SM scalar singlet 
with non-zero vacuum expectation value ({\it vev}) can make the EW 
completely stable due to a tree-level effect on the Higgs quartic coupling 
which may be enough  to guarantees the stability 
at large Higgs field values \cite{Lebedev12,Elias12,Ballesteros15}. 

An appealing scenario in the presence of Higgs portal interactions
is given by a SM singlet scalar field with non-zero {\it vev}
mixed  with the SM Higgs boson,  often called dark Higgs boson. 
The dark Higgs mixing with  the SM Higgs boson 
make possible the direct search of the dark Higgs inflaton at collider experiments.
The mixing guarantee that dark Higgs can be produced in the same channels as the SM Higgs boson 
 if its mass would be the same as that of the dark Higgs boson. 
 Through the same mixing  the dark Higgs boson inherits
the SM Higgs boson couplings to SM fermions via the Yukawa interaction term:
\begin{equation}
L  \supset
\theta \frac{m_{f}} {v}  \phi {\bar f}{f} \,,
\end{equation}
 where: $\phi$ is the dark Higgs field, $\theta$ is its  mixing angle with SM Higgs boson and $m_f$ is the fermion mass. 
 
Dark Higgs bosons can be produced at  LHC  in rare heavy meson decays  (such as K and B mesons). 
They are highly collimated, with characteristic angles $\alpha=M/E$ relative to the parent meson's direction ($M$ is the meson mass and $E$ is the dark Higgs energy).
For $E \sim$ 1 TeV the light dark Higgs decay lengths are  of ${\cal O}$($10^3 \,m$).
Therefore a significant number of dark Higgs bosons can be detected in faraway 
detectors of the LHC experiments \cite{PBC}. Thus,  present and future experimental sensitivity
to the light dark Higgs boson decay crucially depends on its production and decay rates and 
on detector's location and acceptance.

The light dark Higgs boson as inflaton (rather than the Higgs boson) has been first 
implemented in Ref. \cite{Tkchev06}, extending the
$\nu$MSM model \cite{nuMSM1,nuMSM2} to simultaneously explain the cosmological inflation, the DM sterile neutrino masses and the baryon asymmetry of the universe \cite{Tkchev06,Anisimov09}. 
The light dark Higgs inflaton properties has been mostly studied in the frame  of
dark Higgs inflation with non-minimal coupling to gravity \cite{Lerner11,Tenkanen16,Aravind16,Kim17}.
Refs.\cite{Bezrukov10,Bezrukov13a} present a detailed analysis on the possibility to explore this model in the particle physics experiments.\\
This possibility has been also investigated in the frame of low-scale inflation models, such the quartic hilltop model \cite{Bramante16} that predicting a very small value for tensor-to scalar ratio, beyond the sensitivity of the CMB experiments. 
Thus, the dark Higgs searchers  at LHC could experimentally test the low-scale of inflation.

In this paper  we analyse the dark Higgs inflation model
with curvature corrections given by the  kinetic term non-minimally coupled to the Einstein tensor and the coupling to the Gauss-Bonnet (GB) 4-dimensional invariant (hereafter EGB dark Higgs inflation)
and explore the possibility to test its predictions 
by the particle physics experiments.\\
In this model, the non-minimal kinetic coupling to the Einstein tensor causes the inflaton field to roll slower, avoiding the problem of large fields present in chaotic inflation \cite{Granda11}. 
On the other hand, the second-order curvature corrections represented by the scalar field coupled to the GB  term
can increment or suppress (depending on the sign) the tensor-to-scalar ratio
\cite{Jiang13,Kanti15,Odintsov18}. The dynamics of the slow-roll inflation 
by combining both corrections has been proposed in context of the SM Higgs inflation in Refs.\cite{Jimenez19a,Jimenez19b}.\\
The possibility to explore this model by the dark Higgs searchers at LHC
could provide connections between fundamental theories like supergravity and string theories where these couplings are expected to arise, and the Higgs portal interactions.

The paper is organised as follows. In the next section we discuss the dark Higgs inflaton properties. 
In Section~3 we introduce the EGB dark Higgs inflation model.
In Section~4 we analyse the cosmological consistency of the EGB dark Higgs inflation
predictions. Section~5  discuss the 
possibility to test the EGB dark Higgs inflation predictions  by some representative particle physics experiments at LHC. In Section~6 we draw our conclusions.\\
Throughout the paper we consider an homogeneous and isotropic flat background described
by the Friedmann-Robertson-Walker (FRW) metric:
\begin{equation}
\label{FRW}
{\rm d}s^2=g_{\mu,\nu}{\rm d }x_{\mu}{\rm d}x^{\nu}=-{\rm d}t^2+a^2(t)dx^2 \,,
\end{equation}
where $a$ is the cosmological scale factor ($a_0$=1 today).  Also, we use 
the overdot to denote the time derivative and $( ' )$ to denote the derivative with respect to the scalar field. 

\section{Dark Higgs inflaton properties}
\subsection{Dark Higgs inflaton parameters}

We consider the extension of the SM canonical action with the dark Higgs inflaton field, as  introduced in Ref. \cite{Tkchev06}:
\begin{eqnarray} 
\label{S} 
S=\int{{\rm d}^4\,x}\sqrt{-g}\, \left[\frac{\cal R}{2\kappa^2} + \frac{1}{2} (\partial_{\mu} \phi)^2 -V(\phi) \right]\,,
\end{eqnarray}
 where ${\cal R}$ is the Ricci scalar, $\kappa^2 = M^{-2}_{pl}$, $\phi$ is the dark Higgs inflaton field with the potential 
 $V(\phi)$ defined as:
\begin{equation}
\label{DH_V}
V(\phi)=-\frac{1}{2}m_{\phi}^2 \phi^2 +\frac{\beta}{4}\phi^4 +
\lambda \left( {\cal H}^{\dagger} {\cal H} -\frac{\alpha}{\lambda}\phi^2\right)^2 \,.
\end{equation} 
In the above equation $\lambda$ is the SM Higgs field self coupling, 
 $m_{\phi}$ is the dark Higgs mass, $\beta$ is the dark Higgs quartic coupling
and  $\alpha$ is the coupling between the SM Higgs field ${\cal H}$ and the dark Higgs inflaton. 
For $\alpha, \beta \ll \lambda$, inflation is driven along a flat direction of the scalar potential given by:
\begin{equation}
\label{flat}
{\cal H}^{\dagger} {\cal H} \simeq \frac{\alpha}{\lambda} \phi^2 \,.
\end{equation} 
Along this direction  the dark Higgs potential is $V(\phi) = \beta \phi^4/4$ and the coupling constant $\beta$ 
can be fixed from the requirement to obtain the correct the amplitude of the curvature density  perturbations.
This condition leads to $\beta \sim 1.3\times 10^{-13}$ \cite{Lyth99}. \\
The negative sign of the quadratic term in (\ref{DH_V}) ensures that  the scale invariance 
is explicitly broken on the classical level in the inflaton
sector, leading to non-zero {\it vev} for the dark Higgs inflaton after reheating. Then, 
the condition (\ref{flat}) gives rise to EW spontaneous symmetry breaking and the SM Higgs field 
gains non-zero {\it vev } too. 

We remind that the SM Higgs boson mass is given by $m_{H}=\sqrt{2 \lambda}\, v$, where  the SM Higgs {\it vev}  is fixed at $v\equiv(\sqrt{2} G_F)^{1/2}=246.22$ GeV by the Fermi coupling constant 
$G_F$, and the experimentally measured Higgs boson mass is $m_H= 125.10\pm 0.14$ GeV \cite{PDG}. \\
In the gauge base $(\sqrt{2}{\cal H} - {\it v}, \phi)$
the dark Higgs field expectation value, $<\phi>$, its mass $m_{\phi}$ and mixing angle $\theta$, are given by:
\begin{eqnarray}
\label{DH_par}
< \phi > = \frac{m_H}{2 \sqrt{\alpha}  }\,, \hspace{0.5cm} m_{\phi} = m_H \sqrt{ \frac{\beta}{2 \alpha}}\,,
\hspace{0.5cm} \theta=\sqrt{ \frac{2 \alpha}{\lambda}}\,.
\end{eqnarray}
For the purpose of this work we choose $\alpha > \beta/2$, therefore the dark Higgs inflaton is lighter than  the SM Higgs boson, $m_{\phi} < m_H$.
\begin{figure}
\label{fig1}
\centering
\includegraphics[width=7cm,height=7cm]{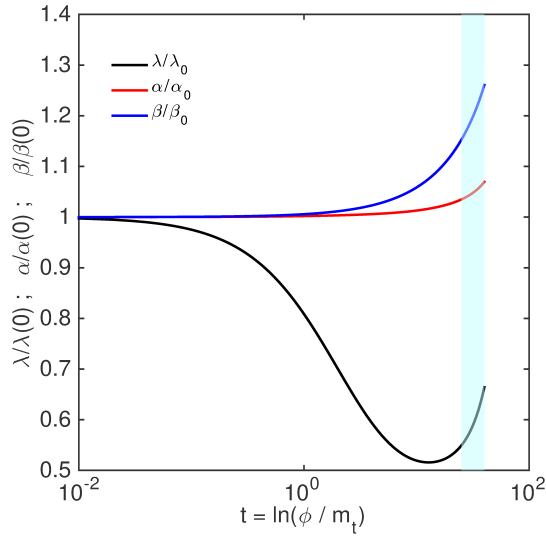}
\caption{The evolution with the scale dependent variable $t={\rm ln}(\phi/m_t)$ of the running of 
$\lambda$, $\beta$ and $\alpha$ coupling constants normalised to their initial values chosen at $t=0$
as: $\lambda(0)=0.129$, $\beta(0)=1.3 \times 10^{-13}$ and  $\alpha(0)=3 \times 10^{-7}$. 
The SM Higgs mass is fixed at $m_H=125.09$ GeV.
The right-hand blue region indicates the slow-roll inflationary regime. \label{Fig1}}.
\end{figure}

The upper bound on the coupling constant $\alpha$ cames from the requirement 
that the quantum corrections do not upset
the flatness of the inflation potential. 
This constrain leads to  $\alpha < 3 \times 10^{-7}$ at the tree level \cite{Bezrukov10} and corresponds 
to the lower bound of the dark Higgs inflaton mass:
\begin{equation}
m_{{\phi}} \geqslant 0.058 \left ( \frac{\beta}{1.3 \times 10^{-13}} \right)^{1/2} \,\, {\rm GeV}.
\end{equation}
The lower bound on $\alpha$ cames from the requirement 
to have an  efficient conversion of the lepton asymmetry to baryon asymmetry
during baryogengesis. This requirement leads to  to $\alpha > \beta \sim 10^{-13}$ \cite{Tkchev06}. 
A stronger lower bound on $\alpha$ is placed by the estimate of the reheating temperature.
For the inflaton particles in thermal equilibrium the reheating temperature is given by \cite{Anisimov09}:
\begin{equation}
\label{Treh}
T_r \simeq  \frac{\zeta(3) \alpha^2} {4 \pi^2} \sqrt {\frac{90}{g_r}} \,{\rm  M_{pl}} \,,
\end{equation}
where $g_r=106.75$ is the SM effective number of relativistic degrees of freedom at reheating and 
 $\zeta(3)=1.202$ is the Reimann zeta function.
The requirement  that $T_r  > 150$ GeV ($T  \simeq 150 $ GeV  is the temperature of the EW symmetry breaking), 
leads to $ \alpha >7.3 \times 10^{-8}$.\\
 For a  non-thermal distribution of the inflaton field the estimate of the reheating temperature 
 is $\sim 10^5 T_r$ \cite{Micha03,Micha04}, leading to $\alpha > 7 \times 10^{-10}$. \\
 These constraints are consistent to the upper bounds of the dark Higgs inflaton mass \cite{Anisimov09}:
\begin{equation}
m_{{\phi}} \leqslant  (0.116 -1.166) \left( \frac{\beta} {1.3 \times 10^{-13} } \right) ^{1/2}\,\, {\rm GeV}\,,
\end{equation}
where the range corresponds to the thermal or non-thermal estimates. 

The above bounds of the inflaton mass may be changed  
if the quantum corrections of the coupling constants are take into account. \\
Figure~\ref{Fig1} presents the evolution with the scale dependent variable $t={\rm ln}(\phi/m_t)$ of the running of 
$\lambda(t)$, $\beta(t)$ and $\alpha(t)$ coupling constants normalised to their initial values 
$\lambda(0)$, $\beta(0)$ and $\alpha(0)$, obtained by the integration the corresponding beta functions  \cite{Degrassi12,Lerner11,Kim17,Aravind16}. \\
As the SM Higgs mass is fixed at $m_H=125.09$ GeV we take $\lambda(0)=0.129\,$.
We also fix  $\alpha(0)=3 \times 10^{-7}$ to the $\alpha$ upper bound and take $\beta(0)=1.3 \times 10^{-13}$. \\
One should note that the correction to $\beta$ from the coupling of the dark Higgs inflaton  to the SM Higgs boson is
$ \delta \beta \sim \alpha^2$ and therefore the evolution of $\beta$ is dominated by  the $\alpha^2$ contribution.
Figure~1 shows that all coupling constants remain positive at the inflationary scale while  the
 flatness of the inflationary potential is preserved.
 
 \subsection{Reheating and horizon crossing}

The reheating proceeds by the energy transfer from the dark Higgs inflaton field to the SM Higgs particles  
through a regime of parametric resonance \cite{Micha03,Micha04}. 
At early stages the entire energy is  in the inflaton zero-mode
and all other modes are absent. The inflaton zero-mode 
oscillations excite the non-zero modes of the inflaton and  of the SM Higgs particles.
This parametric resonance regime ends before a significant
part of the inflaton zero-mode energy is depleted \cite{Anisimov09}.
The reason is the SM Higgs re-scattering process that become important quite early 
because of the large SM Higgs sef-coupling ($\lambda \sim 0.1 $).\\
After the end of the parametric resonance regime, the fluctuations of the inflaton 
field continue to grow exponentially while the energy transferred to the SM Higgs field 
is negligible small. 
The  SM Higgs re-scattering processes bring the inflaton particles in the thermal equilibrium
and the reheating proceeds through  the decay
of the dark Higgs inflaton into the SM Higgs particles.  

The inflationary observables are evaluated at the epoch of the Hubble crossing scale $k_*$ (pivot scale) quantified 
by the number of {\it e}-folds ${\cal N}$ before the end of the inflation.
Therefore, the uncertainties in the determination of $\cal N$ translates into theoretical uncertainties in determination of the inflationary observables \cite{Kinney06,Lerner11}.
Assuming that the ratio of the today entropy per co-moving volume  to that after reheating
is negligible, the main error $\Delta {\cal N}$ in the determination of ${\cal N}$ is given by the uncertainty  in
the determination of the reheating temperature ${T_r}$.
The number of {\it e}-foldings at Hubble crossing scale $k_*$ is related to $T_r$ through:
\begin{equation}
\label{N_TR}
 {\cal N}= {\rm log} \left[ \left( \frac{\rho_r}{\rho_{e} }\right)^{1/4} 
 \left(\frac{g_0 T^3_0}{g_r T^3_r}\right)^{1/3} \left(\frac{k_*}{a_0H_0}\right) \right] \,,
\end{equation} 
where $\rho_r$ and $\rho_e$ refer to the densities at reheating and at end of inflation, 
$T_0$ is the present photon temperature, $H_*$ is the Hubble parameter at  $k_*$, 
$g_r=106.75$ and $g_0=43/11$ are the effective number of relativistic degrees of freedom 
at reheating and at present time. \\
From (\ref{Treh}) and ({\ref{N_TR}) we get  $\Delta {\cal N} \simeq 3$  
corresponding to the uncertainty in determination of $T_r$ for to  a thermal distribution
of the inflaton. This uncertainty is four times higher, $\Delta {\cal N} \simeq 12$, in the  case of non-thermal distribution.

\section{Dark Higgs inflation with curvature corrections}

Closely following \cite{Jimenez19b,Jimenez19a}, in this section we introduce the inflation model 
assuming non-minimal coupling of the dark Higgs
field with the Einstein tensor and to the Gauss-Bonnet  (GB) 4~-~dimensional invariant 
(the EGB dark Higgs inflation model),
derive the background field equations, the slow-roll parameters and evaluate
the primordial power spectra  the the scalar and tensor perturbations.\\
The action of this model is:
\begin{equation}
\label{S_E}
S_E=\int{{\rm d}^4\,x}\sqrt{-g}\, \left[\frac{\cal R}{2\kappa^2} + 
X-V(\phi) +F_1(\phi) G_{\mu \nu}  \partial^{\mu} \phi \partial^{\nu} \phi 
-F_2(\phi) {\cal G} \right]\,,
\end{equation}
where $V(\phi)$ is the dark Higgs potential given in 
(\ref{DH_V}), $F_1(\phi)$ and $F_2(\phi)$ are coupling functions, 
 $G_{\mu \nu}$ is the Einstein tensor, $\cal G$ is the GB 4-dimensional invariant:
\begin{equation}
{\cal G}= {\cal R}^2- 4 {\cal R}_{\mu \nu}{\cal R}^{\mu \nu}+{\cal R}_{\mu \nu \delta \rho} {\cal R}^{\mu \nu \delta \rho}\,.
\end{equation}
The field equations in  a spatially  flat background 
described by the FRW metric (\ref{FRW}) are of the form 
(see Appendix B from \cite{Jimenez19b}) :
\begin{equation}
\label{H2}
H^2 =  \frac{\kappa^2}{3}\left(\frac{1}{2} +V(\phi)+9H^2 F_1{\dot \phi}^2+24H^3 {\dot F_2} \right) \,,   
\end{equation}
\begin{eqnarray}
\label{phi_dot}
{\ddot \phi}+ 3 H {\dot \phi} & + & V^{'}+24 H^2(H^2+{\dot H})F^{'}_2+18 H^3F_1{\dot \phi} \\ \nonumber
& + & 12 H {\dot H}F_1 {\dot \phi}+6*H^2F_1{\ddot \phi}+3 H^2 F^{'}_1 +{\dot \phi}^2 =0 \,.
\end{eqnarray} 
The slow-roll parameters are defined as:
\begin{eqnarray}
\label{slow-roll}
\epsilon_0= - \frac{{\dot H}}{H^2}\,, \hspace{0.2cm}\epsilon_1=\frac{{\dot \epsilon_0}}{H \epsilon_0} \,, 
\hspace{0.2cm} k_0=3F_1{\dot \phi}^2\,, 
\hspace{0.2cm} k_1=\frac{ {\dot k_0 }}{H k_0}\,, \hspace{0.3cm}
\Delta_0=8H{\dot F_2}\,, \hspace{0.2cm} \Delta_1=\frac{ {\dot \Delta_0}}{H\Delta_0}.
\end{eqnarray}
Under the slow-roll conditions ${\ddot \phi} \ll 3 H {\dot \phi} $ and 
$ |\epsilon_0|\,,|\epsilon_1|\,,...|\Delta_1|\, \ll 1$ the potential and field equations  take the 
form:
\begin{eqnarray}
\label{eq_field}
H^2 &\simeq & \frac{\kappa^2}{3}V(\phi) \,\\
3H{\dot \phi}  & \simeq&   -V^{'} -18 H^3F_1{\dot \phi}-24H^4F^{'}_2 \,.
\end{eqnarray}
The number of {\it e}-folds before the end of inflation expressed in terms of the inflaton field is given by:
\begin{equation}
\label{folds}
{\cal N}= \int^{\phi_E}_{\phi_I} \frac{H}{{\dot \phi}} {\rm d} \phi=\int^{\phi_e}_{\phi_{\cal N}} \frac{H^2+6H^4F_1}
{-8H^4F^{'}_2- \frac{1}{3} V^{'}} {\rm d} \phi \,,
\end{equation}
where $\phi_I$ and $\phi_E$ are the values of the  inflaton field at the begging and at the end of inflation.\\
The power spectra of the primordial scalar and tensor perturbations, $\cal {P}_R$ and ${\cal P}_T$, are computed as:
\begin{eqnarray}
\label{PFS}
{\cal P}_R & =&   A_S \frac{H^2}{2 \pi^2} \frac{{\cal G_S}^{1/2}} {{\cal F_S}^{3/2}}\,, 
\hspace{0.1cm}\hspace{0.4cm}A_S=\frac{1}{2}2^{2\mu_S-3} \left |\frac{\Gamma(\mu_S)}{\Gamma(3/2)}\right|^2,
\hspace{0.1cm} \mu^2_S=\frac{9}{4}\left[1+\frac{4}{3}\epsilon_0+\frac{2}{3}
\frac{2 \epsilon_0 \epsilon_1 -\Delta_0 \Delta_1}{2 \epsilon_0-\Delta_0}\right] \nonumber \\
{\cal P}_T& = & 16 A_T \frac{H^2}{2 \pi^2} \frac{{\cal G_T}^{1/2}} {{\cal F_T}^{3/2}}, 
\hspace{0.2cm} A_T=\frac{1}{2} 2^{2 \mu_{T}-3} \left| \frac{\Gamma(\mu_T)}{\Gamma(3/2)} \right|^2, 
\hspace{0.2cm} \mu_T=\frac{3}{2}+\epsilon_0 \,,
\end{eqnarray}
\begin{eqnarray}
{\cal F_S}=c^2_S {\cal G_S}\,,\hspace{0.2cm} 
{\cal G_S} & = & \epsilon_0 -\frac{1}{2}\Delta_0  \,,
\hspace{1.2cm} c^2_S = 1- \frac{ \frac{4}{3} k_0 (\Delta_0+\frac{4}{3} k_0)  
+  \frac{4}{3} k_0\epsilon_0} {2 \epsilon_0-\Delta_0}\,, \nonumber \\
{\cal F_T}=c^2_T {\cal G_T}\,, \hspace{0.2cm}
{\cal G_T} & = &1-\frac{1}{3}k_0-\Delta_0 \,,
\hspace{0.4cm} c^2_T = \frac{3+k_0-3 \Delta_0(\epsilon_0+\Delta_1)}{3-k_0-3\Delta_0} \,,
\end{eqnarray}
where $c_S$ and $c_T$ are  the  
sound speeds of scalar and tensor density perturbations. \\
The spectral index of scalar density perturbations $n_S$ and the tensor-to-scalar ratio expressed
in terms of slow-roll parameters are given by:
\begin{equation}
\label{n_S}
n_S=-2\epsilon_0-\frac{2 \epsilon_0 \epsilon_1-\Delta_0 \Delta_1}{2 \epsilon_0-\Delta_0}\,.
\end{equation}
\begin{equation}
\label{r}
r= 8  \left(\frac {2 \epsilon_0 -\Delta_0} { 1-\frac {1}{3}k_0 - \Delta_0} \right)\,.
\end{equation}
Hereafter we take  $F_1(\phi)$ and $F_2(\phi)$  power-law functions of the form:
\begin{eqnarray}
\label{ST_coupling}
F_1(\phi)=\frac{\gamma}{\phi^4}\,, \hspace{0.5cm}F_2(\phi)=\frac{\eta}{\phi^4}\,,
\end{eqnarray}
where $\gamma$ and $\eta$ are positive constants with the dimension
$[\gamma]=M_{pl}^2$ and  $[\eta]=M^4_{pl}$.\\
For this setup, the first slow-roll parameter $\epsilon_0$  reads as:
\begin{equation}
\label{slow-roll1}   
\epsilon_0=\frac{16}{3}\frac{(3-2\eta\beta)}{(2+\gamma\beta)\phi^2}\,.
\end{equation}
From (\ref{folds}) one gets the number of {\it e}-folds before the end of inflation:
\begin{equation}
\label{folds_1}
{\cal N}=-\frac{3(2+\gamma \beta)}{16(3-2 \eta\beta)}   \phi^2 \bigg\vert^{\phi_E}_{\phi_I}\,.
\end{equation}
The value of the scalar field at  the end of inflation, $\phi_E$, 
is obtained from the requirement $\epsilon_0=1$,  
while (\ref{folds_1}) allows the determination of the inflaton 
field value $\phi_I$  at ${\cal N}${\it e}-folds before the end of inflation:
\begin{eqnarray}
\label{phi_E}
\phi_E= \frac{ 4\sqrt{3-2\eta\beta}} {\sqrt{3(2+\gamma\beta)}}\,, 
\hspace{0.5cm} \phi_I=\sqrt{{\cal N}+1}\phi_E \,.
\end{eqnarray}

\section{Cosmological constraints }

\subsection{Parameterisation and methods} 

The dark Higgs baseline cosmological model is  described 
by the following parameters:
\begin{equation}
\label{base_line}
{\bf P}=\left\{ \Omega_bh^2 \,,\,\Omega_ch^2\,,\,\theta_s\,,\,\tau\,,\,
{\rm log}(10^{10} A_s)\,,\, n_s\,,\, {\cal N}\,,\,\beta\,, \alpha \right\} \,,
\end{equation}
where: $\Omega_bh^2$ is the present baryon energy density, $\Omega_ch^2$
is the present CDM energy density, $\theta_s$ is the ratio of sound horizon to angular diameter distance at decoupling, $\tau$ is the optical depth at reionization, $A_s$ and $n_s$ are the amplitude and the spectral index of the primordial curvature  perturbations,
${\cal N}$ is the number of {\it e}-folds introduced to account for the uncertainty in the determination of the 
reheating temperature, $\beta$ 
is the dark Higgs quartic coupling and $\alpha$ is the dark Higgs - SM Higgs coupling constant. \\
The EGB dark Higgs inflation model extends the dark Higgs baseline model by including the following parameters: 
\begin{equation}
{\bf P}_{\rm EGB}=\left\{ \gamma \beta\,, \eta \beta\right\} \,,
\end{equation}
where the coupling constants $\gamma$ and $\beta$ are defined in  (\ref{ST_coupling}).

We compute the dependence on the scaling variable  $t = {\rm ln} (\phi / m_t)$ 
 of the running of various coupling constants by integrating the corresponding  beta functions:
\begin{eqnarray}
Y(t)=\int^{t}_{0} { {\bf \beta} }_Y {\rm d} t \,, \hspace{1cm} Y=\{g,\, g^{'} \,, g_{s} \,,y_{t}\,,\beta \,,\alpha \} \,,
\end{eqnarray}
where $ g,\, g^{'} \,, g_{s}$ are the gauge couplings, $y_t$ is the Yukawa coupling, $\beta$ 
is the dark Higgs quartic coupling and $\alpha$ is the dark Higgs - SM Higgs coupling  (for the relevant beta functions see Appendix A from \cite{Kim17} and references therein).\\
At $t=0$  the SM Higgs self coupling $\lambda(0)=0.129$ and  the top Yukawa coupling $y_t(0)=0.976$  are fixed 
by the SM Higgs and top quark pol masses \cite{Degrassi12}.
For the  gauge couplings at $t=0$ we take $g^{'}(0)=0.364$, $g(0)=0.64$ and $g_s(0)=1.161$ \cite{Barvinsky09}.
The priors for $\beta(0)$ and $\alpha(0)$  are given in Table~1 (see below).

We modify the standard Boltzmann code \texttt{camb}\footnote{\url{http://camb.info}} \cite{camb} to 
calculate the primordial power spectra of scalar ${\cal P}_R(k)$ and tensor ${\cal P}_T(k)$ density perturbations  
for the dark Higgs inflation model with curvature  corrections presented in the previous section.
The code evolves the coupled dark Higgs field equations (\ref{H2}) and (\ref{phi_dot}) 
with respect to the conformal time for wave numbers in the range $5 \times 10^6$~-~$5$ Mpc$^{-1}$ and evaluate the 
RG corrections to the coupling constants as presented before.
The value of the inflaton field $\phi_I$ and $\phi_E$ at the beginning and at the end of inflation 
are obtained from (\ref{phi_E}). 
The primordial power spectra ${\cal P}_R(k)$ and  ${\cal P}_T(k)$ are 
obtained from (\ref{PFS}) with the slow-roll parameters defined in (\ref{slow-roll}).\\
The scalar spectral index of the curvature perturbations $n_S$
and the ratio of tensor-to-scalar amplitudes $r$ are then evaluated 
at the pivot scale $k_*=0.002$Mpc$^{-1}$ as:
\begin{eqnarray}
 n_S=\frac{{\rm d \, ln}{\cal P}_{R}(k)}{ {\rm d\, ln}  k   }\bigg\vert _{k_*}\,,
 \hspace{0.5cm} r=\frac{{\cal P}_T(k)}{ {\cal P}_R(k)} \bigg\vert_{k_*} \,.
\end{eqnarray} 

The extraction of parameters from the cosmological dataset
is based on Monte-Carlo Markov Chains (MCMC) technique.  
We modify the publicly available version of the package 
\texttt{CosmoMC}\footnote{\url{http://cosmologist.info/cosmomc/}} 
\cite{cosmomc} to sample from the space of  dark Higgs inflation model 
parameters and generate estimates of their posterior mean and  confidence intervals.\\
We made some test runs to optimise the parameters prior intervals and sampling. 
The final run is based on 120 independent channels, reaching the convergence criterion 
$(R-1) \simeq 0.01$. The $(R-1)$ criterion is defined as 
the ratio between the variance of the means and the mean of variances for the second half of chains \cite{cosmomc}.\\
We assume a flat universe and uniform priors for all parameters adopted in the analysis in the intervals listed in Table~1.
The Hubble expansion rate $H_0$ 
is a derived parameter in our analysis. We constrained $H_0$ values  to reject the extreme models.

For the cosmological analysis we use the CMB temperature (TT), polarization (EE,TE) and lensing angular power spectra from {\sc Planck} {\texttt 2018} release \cite{Planck_params} and the likelihood codes corresponding to different multipole ranges \cite{Planck_likes}\footnote{\url{http://pla.esac.esa.int/pla/cosmology}}. 
The {\sc Planck} data currently provide the best characterisation of the primordial density perturbations \cite{Planck_infl}, constraining the cosmological parameters  at the sub-percent level \cite{Planck_params}.

We use the following combinations of TT, TE, EE and lensing {\sc Planck} likelihoods  \cite{Planck_infl}:\\
(i) Planck TT+lowE: the combination of  high-{\it l} TT likelihood at multipoles {\it l} $\ge$ 30, the Commander likelihood
 for low-{\it l} temperature-only and the SimAll low-{\it l}
EE likelihood in the range 2 $ <$ {\it l} $<$ 29;
(ii) {\sc Planck} TE and Planck EE: the combination  of TE and EE likelihoods at  {\it l} $\ge$ 30; 
(iii) {\sc Planck} TT,TE,EE+lowE: the combination of Commander likelihood using TT, TE, and EE spectra
at $\it l$ $\ge $30, the low-{\it l} temperature, and the low-{\it} SimAll EE likelihood; 
 (iv) {\sc Planck} TT,TE,EE+lowP: the combination of the likelihoods using TT, TE, and
EE spectra at  {\it l} $>$ 30;
(v) {\sc Planck} high-{\it l} and Planck low-{\it l} polarization:  the Plik likelihood;
(vi) {\sc Planck} CMB lensing:  the CMB lensing likelihood \cite{Planck_lens}  for lensing multipoles 8 $<$ {\it l} $<$ 400.

We also consider the measurement of the CMB B-mode polarization angular power spectrum by
the BICEP2/Keck Array collaboration \cite{BK15}. 
The BK15  likelihood B-mode  polarization  only
leads to an upper limit of  tensor-to-scalar ratio amplitudes 
$r <$ 0.07 (95\% CL) \cite{BK15}.\\\\
We will refer to the combination of these datasets as  {\sc Planck} TT,TE,EE+lowE+lensing+BK15.

\begin{table}
\caption{ Priors and constraints on EGB dark Higgs inflation model parameters adopted in the analysis. 
All priors are uniform in the listed intervals. We assume a flat universe.}
\begin{center}
\begin{tabular}{|l|c|}
\hline 
                 Parameter&Prior  \\
\hline
$\Omega_bh^2$&     [0.005,\,0.1]       \\
$\Omega_ch^2$&     [0.001,\,0.5 ]       \\
$100\theta_s$ & [0.5,\,10] \\
$\tau$& [0.01,\,0.9] \\
${\rm log}(10^{10} A_s)$ & [2.5,\, 5]\\
$n_s$& [0.5,\,1.5]\\
${\cal N}$ &[54,\,64]\\
$\alpha \times 10^{7}$& [0.007,\,3]                 \\
$\beta \times 10^{13}$& [1,\,5]        \\
$\gamma \beta$& [0,\,3] \\
$\eta \beta$ & [0\,,3] \\
\hline
$H_0({\rm km\,s}^{-1}{\rm Mpc}^{-1})$& [20,\,100] \\
\hline
\end{tabular}
\end{center}
\end{table}

\subsection{Analysis}

Left panel from Figure~\ref{Fig2} presents the marginalised likelihood probability distributions
of the inflationary parameters, $A_s$, $n_s$, $r$ and ${\cal N}$
from the fit of the EGB dark Higgs inflation model 
with the  {\sc Planck} TT,TE,EE+lowE+lensing+BK15 dataset.
These predictions are computed 
 at pivot scale $k_*$=0.002 Mpc$^{-1}$ and include 
the uncertainty in the number of e-folds. 
For comparison, we also show the corresponding 65\% and 95\%  limits 
from the fit of $\Lambda$CDM model with the same dataset \cite{Planck_infl}. 
The right panel from the same figure presents  the joint confidence 
regions (68\% and 95\% CL)  of $n_s$ and $r$.\\
The mean values and the errors for all parameters are presented in  Table~2.\\
We find that the EGB dark Higgs inflation model is  strongly favoured by the {\sc Planck}+BK15 data \cite{Planck_infl}.

We test the consistency of the EGB dark Higgs inflation model predictions for the mean values of $\gamma \beta$, $\eta \beta$ and ${\cal N}$ given in Table~2. 
 From (\ref{phi_E}) we evaluate the dark Higgs field at the beginnig of inflation,  $\phi_I=10.36 M_{pl}$. 
The slow-roll parameters  at $\phi_I$ defined in (\ref{slow-roll1}) are given by:
\begin{eqnarray}
\label{slow-roll-fit}
\epsilon_0 & = & \epsilon_1=0.0163, \hspace{0.3cm} k_0=0.0017\,,\hspace{0.3cm}k_1= 0.0165\,,
\hspace{0.3cm}\Delta_0=0.0249\,,\hspace{0.3cm} \Delta_1=0.0165 \,,
\end{eqnarray}
while the  tensor-to-scalar ratio (\ref{r}) and the amplitude of the curvature perturbations (\ref{PFS})
at $\phi_I$ are:
\begin{eqnarray}
 r=0.065\,, \hspace{0.5cm}{\cal P}_R= 1.472 \times 10^{-9}\,.
\end{eqnarray}
The inflation potential (\ref{infl_scale}) at $\phi_I$ is obtained as:
\begin{eqnarray}
\label{V_I}
V^{1/4}(\phi_I)= \left (\frac{3 \pi^2 {\cal P}_R}{2}r  \right)^{1/4} M_{pl}\simeq 6.11 \times 10^{-3} M_{pl} \simeq 1.46 \times 10^{16} {\rm GeV}\,.
\end{eqnarray}
This constraint applied to the dark Higgs potential $V=\beta \phi_I^4/4$ leads  to quartic coupling 
$\beta < 3.38 \times 10^{-13}$, value  consistent with the inflationary observables and with the dark Higgs parameters. 
 Under the slow-roll conditions, we get from ({\ref{V_I}}) the Hubble parameter at $\phi_I$:
 \begin{eqnarray}
 \label{H_I}
H(\phi_I) \simeq 1.51 \times 10^{-5}M_{pl} \simeq  3.63 \times 10^{13}{\rm GeV}\,.
 \end{eqnarray}
From (\ref{H_I}) it follows that the curvature scale at $\phi_I$ satisfy the condition 
$R\simeq12H^2 \ll M^2_{pl}$ and therefore the unitarity bound of dark Higgs inflation model with curvature couplings is not exceeded. 
\begin{figure}
\centering
\includegraphics[width=13cm,height=7cm]{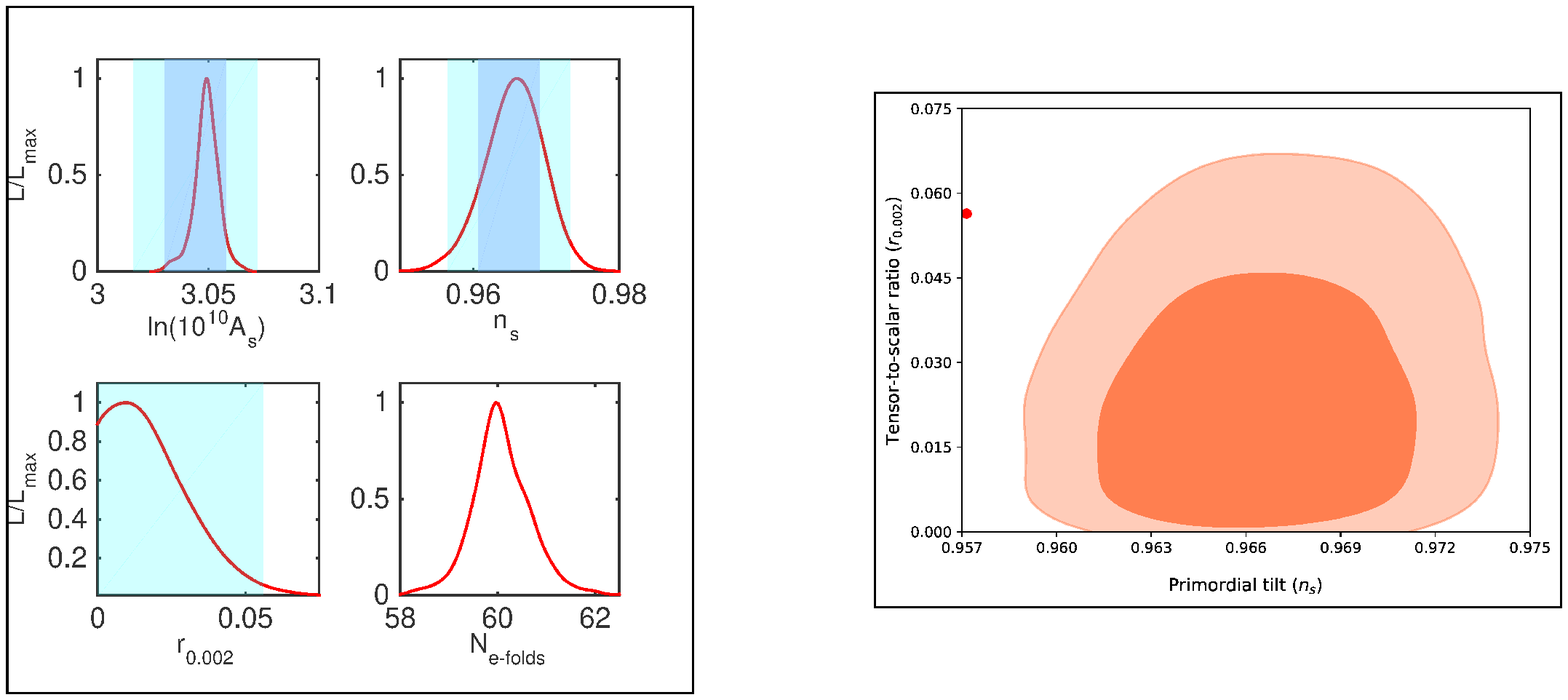}
\caption{{\it Left}: Marginalised likelihood probability distributions of the main inflationary parameters from the fit of the EGB dark Higgs inflation model with the  {\sc Planck} TT,TE,EE+lowE+lensing+BK15 dataset.
The distributions are obtained at  $k_*$=0.002 Mpc$^{-1}$ and include 
the uncertainty in the number of e-folds. 
For comparison we also show the corresponding 65\% (dark blue) and 95\% (light blue) limits 
from the fit of $\Lambda$CDM model with the same dataset \cite{Planck_infl}.
{\it Right}: Marginalised joint 68\% and 95\% CL regions for $n_s$ and $r$  distributions presented in the  left panel.\label{Fig2}}
\end{figure}
\begin{figure}
\centering
\includegraphics[width=13cm,height=7cm]{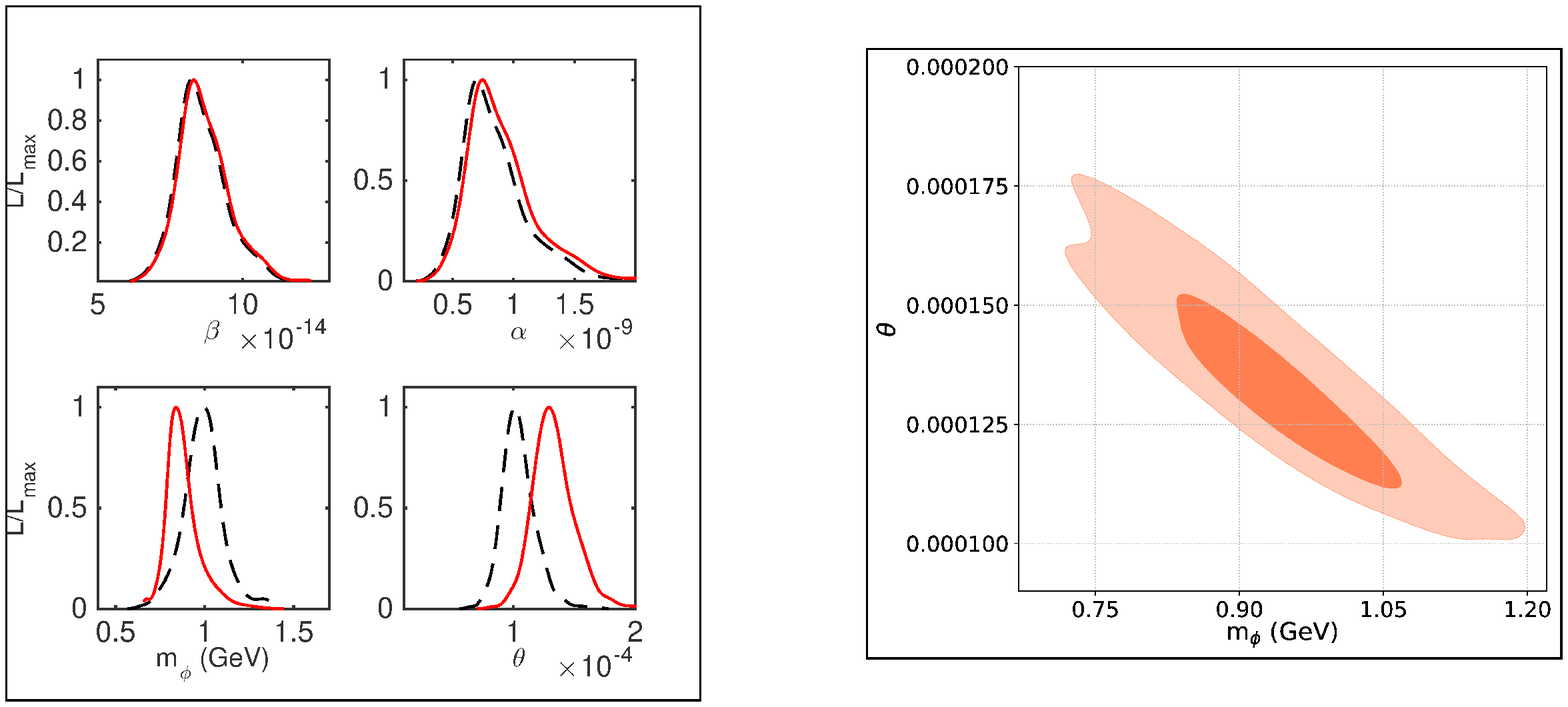}
\caption{{\it Left}: Marginalised likelihood probability distributions of the dark Higgs parameters 
from the fit of the EGB dark Higgs inflation model with the  {\sc Planck} TT,TE,EE+lowE+lensing+BK15 dataset, obtained with (red) and without (dashed black) quantum corrections.
{\it Right}:  Marginalised joint 68\% and 95\% CL regions for $m_{\phi}$ and $\theta$ 
obtained for EGB dark Higgs inflation model.\label{Fig3}} 
\end{figure}

Left panel from Figure~\ref{Fig3} presents the likelihood probability distributions of the dark Higgs parameters $\beta$, 
$\alpha$, $m_{\phi}$ and $\theta$ obtained from the fit of the EGB dark Higgs inflation model with the  {\sc Planck} TT,TE,EE+lowE+lensing+BK15 dataset. These predictions are computed 
 at $k_*$=0.002 Mpc$^{-1}$ and  include  the quantum corrections of the coupling constants. 
 The mean values and the errors of these parameters  are given in Table~2. \\
For comparison  we plot the same distributions without quantum corrections.

The dark Higgs mass $m_\phi$ and mixing angle $\theta$ are derived parameters in our analysis 
and are obtained from (\ref{DH_par}).
The predictions of  the EGB dark Higgs inflation model for the joint confidence 
regions (68\% and 95\% CL)  of $m_{\phi}$ and $\theta$  are shown in the right panel of Figure~\ref{Fig3}. \\
We find  for   the dark Higgs - SM Higgs coupling, 
$ 7 \times 10^{-10}< \alpha < 3 \times 10^{-8}$.
These bounds are in agreement with the estimate of the reheating temperature
 for a non-thermal distribution of the inflaton field \cite{Anisimov09}.\\
The bounds on the dark Higgs mass and mixing angle are found to be:
\begin{eqnarray}
\label{mass_range}
0.49 \,\, {\rm GeV} & < & m_{\phi} <  1.43\,\, {\rm GeV}\,, \hspace{1cm} (95\%\,\,{\rm CL}) \\
4.48 \times 10^{-5} & < & \theta <  1.88 \times 10^{-4} \,.
\end{eqnarray}

\section{Dark Higgs inflaton at LHC experiments}

\subsection{Dark Higgs inflaton decay inside detector}
\begin{figure}
\centering
\includegraphics[width=7cm,height=7cm]{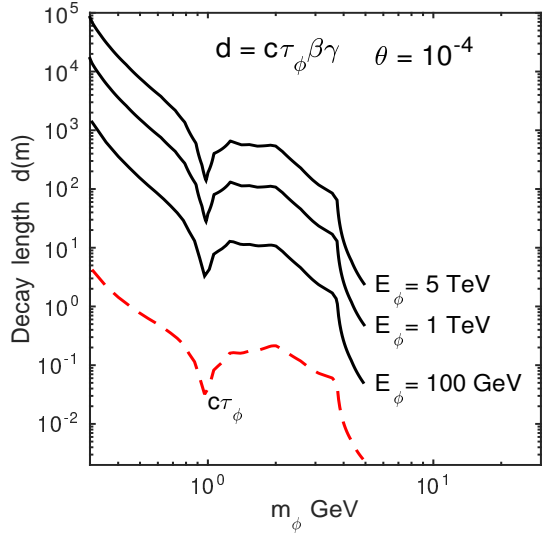}
\caption{The evolution with $m_{\phi }$ of  dark Higgs inflaton decay length,
$d=c \tau_{\phi} \gamma \beta$, for various  dark Higgs energies $E_{\phi}$ and $\theta=10^{-4}$.  \label{Fig4} }
\end{figure}

 
The dark Higgs decay widths are suppressed by $\theta^2$ relative to those of the SM Higgs boson
if it would have the some mass as the dark Higgs. 
For $m_{\phi} < 2m_{ \pi}$ the inflaton mostly decays  in $e^{+} e^{-}$, $\mu^{+}\mu^{-}$  and 
$\tau^{+}\tau^{-}$ with decay width given by:
\begin{eqnarray}
\Gamma (\phi \rightarrow {\bar l} l)= G_F\frac{m^2_{l} m_{\phi}} {4 \sqrt{2} \pi } \beta^3_l \theta^2
\hspace{0.5cm} (l= e\,,\mu\,,\tau)\,,
\end{eqnarray}
where $G_F$ is the Fermi constant and $\beta_l= \sqrt { 1- m^2_l/ m^2_{\phi} }  $ is the lepton velocity.\\
For inflaton masses in the range  $2 m_{\pi} < m_{\phi} < 2.5$ GeV 
the dominant decay modes are to $\pi^{+}\pi^{-}$, $k^{+}k^{-}$ and other hadrons.\\
The dark Higgs hadronic decay modes 
suffers from theoretical uncertainties since the chiral expansion breaks down above 
$2m_{\pi}$ while the perturbative QCD calculation
are reliable for masses of  few GeV \cite{Clarke14,Winkler19}.\\
For the inflaton mass range (\ref{mass_range}) we adopt the numerical 
results from \cite{Winkler19} that uses the  dispersive analysis  
for $2 m_{\pi} < m_{\phi}<  1.3$ GeV \cite{Grinstein88},
the perturbative spectator model \cite{Guinon00,Keen09} for $m_{\phi}\,>\,$ 2GeV and 
 interpolate between these two for 1.3 GeV$< m_{\phi}< $2 GeV. \\
Figure~\ref{Fig4} presents the dependence on $E_{\phi}$ of the dark Higgs decay length:
\begin{eqnarray}
d=c \tau_{\phi} \gamma \beta \,
\end{eqnarray}
where $\tau_{\phi}=1/\Gamma(\phi \rightarrow ll,hh)$ is the dark Higgs lifetime,
$\Gamma(\phi \rightarrow ll,hh)$ is the decay width scaled with $\theta^2$,  $\gamma=E_{\phi}/m_{\phi}$ and $\beta=\sqrt{1-1/\gamma^2}$. The decay length scales as $d \sim E_{\phi}$ for large $E_{\phi}$. For $E_{\phi}\sim {\cal O} (10^3)$ GeV the decay lengths are d$\sim {\cal O} (1)$ km
and therefore a significant  number of dark Higgs inflatons can decay within 
the detector volume.

To determine the number of dark Higgs inflatons that decay inside the detector volume,  we must specify
the size, shape, and location of the detector relative to the LHC collider interaction point (IP). \\
We consider two representative  experiments, FASER (the ForwArd Search ExpeRiment) \cite{Faser18} and 
MAPP-1(the MoEDAL  Apparatus for Penetrating Particles) (\cite{Pinfold19}).\\
FASER  detectors  have cylindrical shape and are centred on the LHC  beam collision 
axis. The detectors have the radius  R and 
the depth $\Delta= L_{max}-L_{min}$, where $L_{max}$ and  $L_{min}$ are the distances from the 
IP to the far and near edge of detectors along the beam axis. The location of FASER detectors is:
\begin{eqnarray}
{\rm FASER\,\,\,far\,\,\,location}&:& \hspace{0.5cm}L_{max}=400\, m\,, \,\Delta=10\,m, \,R=1\,m \,,\\
{\rm FASER\,\,\,near \,\,\,location}&:& \hspace{0.5cm}L_{max}=150\,m\,,\, \Delta=5\,m\,,\, R=4\,{\rm cm}\,. 
\end{eqnarray}
MAPP detector is a parallelepiped at approximately  $5^0$ from  the beam collision axis with the following location:
\begin{eqnarray}
\label{prob}
\hspace{0.2cm}{\rm MAPP-1}: \hspace{0.5cm}        L_{max}=55\,m\,, \Delta=3\,m\,, H=1\,{\rm m} \,,
\end{eqnarray}
where H is the is the parallelepiped height. \\
The probability of the dark Higgs boson  to decay inside the detector volume is given by:
\begin{equation}
\label{prob}
{\cal P}^{det}(E_{\phi},\theta_{\alpha})=\left( e^{-L_{min}/d}-e^{-L_{max}/d} \right)\Theta(R, \tan(\theta_{\alpha}) L_{max})\,
\end{equation}
where $E_{\phi}$ is the dark Higgs energy, $d$ is its decay length, $\theta_{\alpha}$ is the angular acceptance of the detector, $\tan(\theta_{\alpha})=R/L_{max}$,   
and $\Theta$ is the Heaviside step function.
For MAPP-1 we take $R=H\pi^{-1/2}$ in (\ref{prob}) to conserve the effective acceptance area.\\
In Figure~\ref{Fig5} we present the dependence on $E_{\phi}$ of the normalised detection probability 
corresponding to the different experimental configurations obtained for 
 cosmological best fit  solution for $m_{\phi}$ and $\theta$.
The figure shows that the experimental configurations are sensitive to 
complementar ranges of the dark Higgs energy. 
\begin{figure}
\centering
\includegraphics[width=8cm,height=5cm]{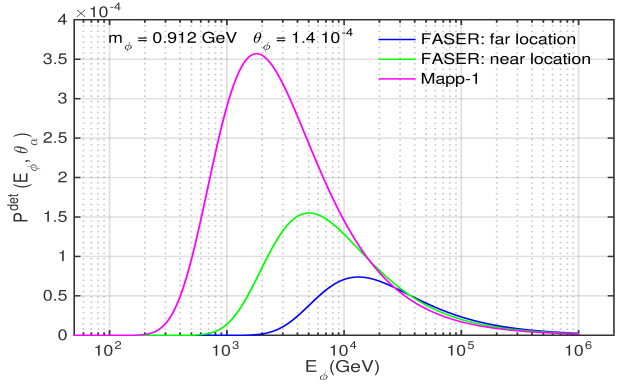}
\caption{ The dependence on the dark Higgs energy $E_{\phi}$ of the normalised detection probability 
corresponding to different experimental configurations obtained for 
 cosmological best fit  solution for $m_{\phi}$ and $\theta$.\label{Fig5}}
\end{figure}

\subsection{Dark Higgs inflaton production at LHC}

The dark Higgs bosons are mainly produced  through K and B meson decays.
As $m_{\phi} > m_K$ ($m_K=0.494$ GeV) for the inflaton mass range (\ref{mass_range}),
in the following we consider the dark Higgs production only through the B-meson decay. 
The B-meson branching fraction is given by \cite{Faser18}:
\begin{equation}
Br(B\rightarrow X_{s} \phi) =5.7 \left(1-\frac{m^2_{\phi}}{m^2_B} \right)^2 \theta^2\,,
\end{equation}
where $X_s$ denotes any strange hadronic state and $m_B$ is the B-meson mass ($m_B=5.28$ GeV).
\begin{figure}
\centering
\includegraphics[width=9cm,height=9cm]{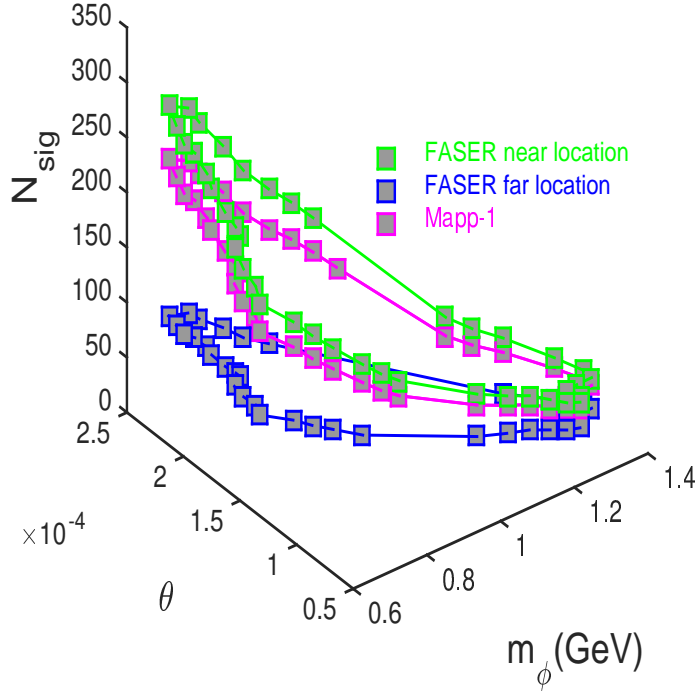}
\caption{LHC experiments reach for dark Higgs inflaton 
in the cosmological confidence region (\ref{mass_range}) for 
an integrated luminosity of 3ab$^{-1}$ at 13 TeV LHC 
assuming 100\% detection efficiency.\label{Fig6}  }
\end{figure}

The dark Higgs production cross section at LHC energies can be estimated as  \cite{Bezrukov10}:
\begin{eqnarray}
\frac{\sigma_{\phi} }{ \sigma_{inel}} =M_{pp} Br(B\rightarrow X_{s} \phi) \,,
\end{eqnarray}
where $M_{pp}$ is the proton multiplicity  and $\sigma_{inel}$ is the $pp$ inelastic cross section.

\subsection{LHC experiments reach for dark Higgs inflaton}

The total number of dark Higgs bosons that decay inside detector are then given by:
\begin{equation}
N_{sig} (m_{\phi},\theta)=N_{inel} \frac{\sigma_{\phi}} {\sigma_{inel}}\,
Br(\phi \rightarrow KK) \,Br(\phi \rightarrow \pi\pi) 
\int {\cal P}^{det} 
(E_{\phi},\theta_{\alpha})
 {\rm d\,} \theta_{\alpha} {\rm d}\,E_{\phi} \,,
\end{equation}
where $N_{inel}$ is the total number of inelastic  $pp$ scatering events.\\
 Throughout we assume an integrated luminosity of 3 a$b^{-1}$ at the 13 TeV LHC, implying 
 $N_{inel} \simeq1.1 \times 10^{16}$. We also take  $\sigma_{inel}$(13 TeV) $\simeq$ 75 mb and $M_{pp}$(13 TeV) $\simeq$ 66 \cite{PDG}.
 
Figure~6 shows the predicted number of dark Higgs inflaton decays in 
the cosmological confidence region (\ref{mass_range}) obtained for the experimental configurations  discussed for an integrated luminosity of 3ab$^{-1}$ at 13 TeV LHC assuming 100\% detection efficiency. \\
In our computation  we take  the dark Higgs energy in the range 
100 GeV $< E_{\phi} <  10{^6}$ GeV, imposed by
the requirement that the dark Higgs inflaton propagate to the detector locations, as 
shown in Figure~\ref{Fig5}. \label{Fig6}}\\
For comparison, in Figure~\ref{Fig7} we present the FASER reach \cite{Faser18} and the MAPP-1 reach \cite{Pinfold19} for the dark Higgs boson for an integrated luminosity of 3ab$^{-1}$ at 13 TeV LHC.
The cosmological dark Higgs inflaton confidence region (\ref{mass_range}) is also shown. 
\begin{figure}
\centering
\includegraphics[width=7cm,height=7cm]{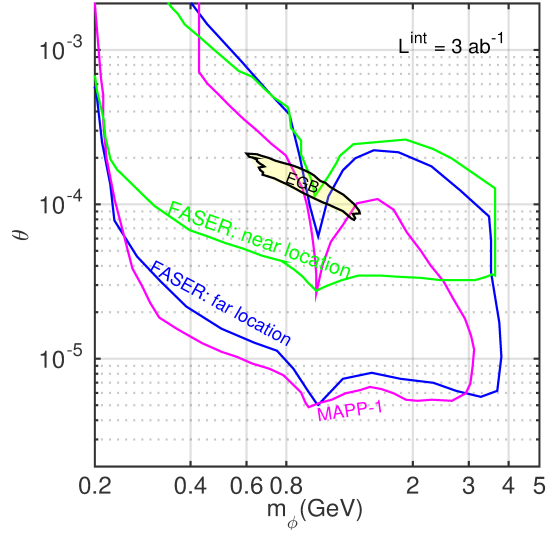}
\caption{ FASER far location, FASER near location  and MAPP-1 reach for dark Higgs boson
for an integrated luminosity of 3ab$^{-1}$ at13 TeV LHC.
The cosmological dark Higgs inflaton confidence region (\ref{mass_range}) is also shown.
\label{Fig7}}
\end{figure}
\begin{table}
\caption{The mean values and the absolute errors of the main parameters obtained from the fit
of the  EGB dark Higgs inflation model with {\sc Planck} TT,TE,EE+lowE+lensing+BK15 dataset. The errors are
quoted at 68\% CL. 
The upper limits are quoted at 95\% CL.
The first group of parameters are the
base cosmological parameters sampled in the Monte-Carlo Markov Chains analysis with uniform
priors.The others are derived parameters.}
\begin{center}
\begin{tabular}{|l|c|}
\hline 
Parameter                  &     \\
\hline
$\Omega_b h^{2}$& 0.0223 $\pm$ 0.0002\\
$\Omega_c h^{2}$  & 0.1194 $\pm$  0.0011 \\
$\theta_s$ & 1.0410 $\pm$ 0.0004 \\
$\tau$	    &0.050 $\pm$ 0.009 \\
${\rm ln}(10^{10}A_s)$     &3.050 $\pm$ 0.008 \\
$n_s$   & 0.967 $\pm$ 0.0044 \\
$r_{0.002}$& $<$ 0.059 \\
${\cal N}$& 59.4 $\pm$ 1.210 \\
$10^{13} \times\beta$&0.892 $\pm$ 0.051    \\ 
$10^{9}\times\alpha$  & 1.021 $\pm$0.219           \\
$\gamma \beta$    & 0.218 $\pm$ 0.015       \\
$\eta \beta$& 1.129 $\pm$ 0.067        \\
\hline
$H_0({\rm km\,s}^{-1}{\rm Mpc}^{-1})$ &67.729 $\pm$ 0.641 \\
$m_{\phi}$ (GeV)  &   0.919 $\pm$ 0.211      \\ 
$10^4 \times \theta$    & 1.492 $\pm$ 0.045      \\ 
\hline 
\end{tabular}
\end{center}
\end{table}
\section{Conclusion}

In this paper we analyse the dark Higgs inflation model with curvature corrections
given by the kinetic term non-minimally coupled to the Einstein tensor and the coupling to
the Gauss-Bonnet (GB) 4-dimensional invariant (EGB dark Higgs inflation)  and
explore the possibility to test its predictions by particle physics experiments at LHC.\\
The dynamics of the slow-roll inflation with curvature corrctions
has been proposed in context of the SM Higgs inflation in Refs \cite{Jimenez19a,Jimenez19b}.

We show that the EGB dark Higgs inflation model 
is  strongly favoured by  {\sc Planck}+BK15 data \cite{Planck_infl}.
The cosmological predictions of this model for dark Higgs inflaton mass 
$m_{\phi}$ and mixing angle $\theta$, including the RG quantum corrections of dark Higgs coupling constants and the uncertainty in estimation of the reheating temperature, 
are found to be:
\begin{eqnarray}
\label{mass_range_fin}
\label{mass_range_RG}
0.49 \,\, {\rm GeV} & < & m_{\phi} <  1.34\,\, {\rm GeV}\,, \hspace{1cm} (95\%\,\,{\rm CL}) \nonumber \\
4.48 \times 10^{-5} & < & \theta <  1.88 \times 10^{-4} \nonumber \,.
\end{eqnarray}
The consistency test of the EGB dark Higgs inflation model predictions leads to 
a lower bound of dark Higgs inflaton quartic coupling $\beta < $ 3.38 $\times 10^{-13}$, 
value consistent with the inflationary observables and with the dark Higgs parameters.\\
We find the dark Higgs inflaton - SM Higgs boson coupling constant
$\alpha > 7\times 10^{-10}$, in agreement with the estimate of the reheating temperature for a non-thermal distribution of the inflaton field \cite{Anisimov09}.

We evaluate the FASER and MAPP-1 experiments reach for dark Higgs inflaton parameters
$m_{\phi}$ and $\theta$ in the 95\% CL cosmological confidence region, for an integrated luminosity of 3ab$^{-1}$ at 13 TeV LHC assuming 100\% detection efficiency. \\
We conclude that the  dark Higgs inflation model with curvature corrections 
is a compelling inflation scenario based on particle physics theory  
favoured by the present cosmological measurements that leaves imprints in the dark Higgs boson searchers at LHC.

\begin{acknowledgments}
The author would like to thank to Vlad Popa for helpful discussions 
and acknowledge the use of the GRID system computing facility at the Institute of Space Science.\\
This work  was partially supported by ESA/PRODEX Contract 4000124902.
\end{acknowledgments}

\end{document}